\documentclass[conference,a4paper,twocolumn]{IEEEtran}

\IEEEoverridecommandlockouts

\usepackage{cite}

\ifCLASSINFOpdf
\else
\fi
\usepackage[cmex10]{amsmath}
\usepackage[ruled,vlined,linesnumbered]{algorithm2e}  
\SetAlgoSkip{}

  \usepackage[caption=false,font=footnotesize]{subfig}
\usepackage{stfloats}
\usepackage{graphicx}
\usepackage{amssymb}
\usepackage{hyperref}
\usepackage{xcolor}
\usepackage[T1]{fontenc}
\usepackage[utf8]{inputenc}
\usepackage{balance}

\usepackage{pgfplots}
\pgfplotsset{compat=newest}
\newlength\figurewidth

\usepackage{pifont}

\newcommand{\qm}[1]{``#1''}  

\newcommand{\oznaceniMetody}[1]{{#1}}

\newcommand{\hankelI}{$\textup{IPMS}_{\Gammaset}$}
\newcommand{\hankelHI}{$\textup{IPMS}_{H\cap\Gammaset}$}
\newcommand{\hankelHIxfade}{$\textup{IPMS}_\textup{xfade}$}
\newcommand{\hankelHIola}{$\textup{IPMS}_\textup{OLA}$}

\def\RR{\mathbb{R}}

\def\x{\vect{x}}

\def\y{\vect{y}}
\def\z{\vect{z}}

\def\d{\vect{d}}

\def\X{\matr{X}}
\def\Y{\matr{Y}}
\def\Z{\matr{Z}}
\def\D{\matr{D}}
\def\I{\matr{I}}

\newcommand{\vect}[1]{\mathbf{#1}} 
\newcommand{\matr}[1]{\mathbf{#1}} 

\newcommand{\argmin}{\mathop{\operatorname{arg\,min}}}

\newcommand{\rank}{\mathop{\operatorname{rank}}}

\newcommand{\proj}{\mathrm{proj}}

\def\Gammaset{\varGamma}

\definecolor{matlabCommentGreen}{RGB}{60,118,61}

\usepackage{listings}

\makeatletter
\def\footnoterule{\kern-3\p@
  \hrule \@width 2in \kern 2.6\p@} 
\makeatother

\begin{document}



\title{Multiple Hankel matrix rank minimization\\ for audio inpainting}


\author{\IEEEauthorblockN{
Pavel Záviška, Pavel Rajmic, and Ondřej Mokrý
}
\IEEEauthorblockA{
Brno University of Technology\\
Faculty of Electrical Engineering and Communications, Dept.\ of Telecommunications\\
Technická 12, 616\,00, Brno, Czech Republic\\
Email: pavel.zaviska@vut.cz, pavel.rajmic@vut.cz, ondrej.mokry@vut.cz
}
\thanks{
The authors would like to thank R.\,Sasaki for providing the Matlab implementation related to the paper \cite{Sasaki2018:Declipping.multiple.matrix.rank.minimization}
and S.\,Ghanbari for preliminary research on the Hankel matrix rank minimization-based audio declipping.

This work was supported by projects 20-29009S and 23-07294S of the Czech Science Foundation (GAČR).
}
}


%


\maketitle


\begin{abstract}
Sasaki et al.\ (2018) presented an efficient audio declipping algorithm,
based on the properties of Hankel-structure matrices constructed from time-domain signal blocks.
We adapt their approach to solving the audio inpainting problem, where samples are missing in the signal.
We analyze the algorithm and provide modifications, some of them leading to an improved performance.
Overall, it turns out that the new algorithms perform reasonably well for speech signals but they are not competitive in the case of music signals.
\end{abstract}



%
\IEEEpeerreviewmaketitle


\section{Introduction}
Audio declipping and audio inpainting are two closely related inverse problems.
In the inpainting case, some audio samples are missing
and the need for a~means of reconstruction naturally arises.
A~number of successful algorithms have been proposed, based on different signal models.
These include assumptions on autoregressivity of audio waveform
\cite{javevr86,Etter1996:Interpolation_AR}
or its smoothness
\cite{Selesnick2013:LeastSquares},
sparsity of the time-frequency audio representation 
\cite{Adler2012:Audio.inpainting,%
MokryRajmic2020:Inpainting.revisited,%
MokryZaviskaRajmicVesely2019:SPAIN,%
TaubockRajbamshiBalasz2021:SPAINMOD}
and low-rank expansions of the spectrogram
\cite{Mokry2022:Audio.inpainting.NMF}.
Other methods rely on copying non-local audio information into the gap
\cite{Perraudin2018:Similarity.Graphs,Bahat_2015:Self.content.based.audio.inpaint}.
A~special class of methods uses deep neural networks to learn the signal reconstruction
\cite{Marafioti2019:DNN.inpainting,Marafioti2021:GACELA}.

The case of audio declipping differs from inpainting solely by additional amplitude-based constraints stemming from the clipping process
\cite{ZaviskaRajmicOzerovRencker2021:Declipping.Survey}.
Thus, if the respective model allows it, declipping methods can be basically identical with their inpainting variants,
with only additional requirements on the feasible set
\cite{SiedenburgKowalskiDoerfler2014:Audio.declip.social.sparsity,%
Kitic2015:Sparsity.cosparsity.declipping,%
ZaviskaRajmicSchimmel2019:Psychoacoustics.l1.declipping,%
GaultierKiticGribonvalBertin:Declipping2021,%
Zaviska2022:Analysis.social.sparsity}.
In some cases, however, such a~modification is not possible and therefore there exist a~variety of algorithms designed specifically for declipping
\cite{Sasaki2018:Declipping.multiple.matrix.rank.minimization,%
BilenOzerovPerez2015:declipping.via.NMF,%
HarvillaStern2015:Parameter.estimation.declipping.in.noise}.

In the present paper,
the fundamental assumption is that a~segment of audio can be effectively approximated using an autoregressive (AR) process.
This assumption is nevertheless not utilized directly by modeling the latent 
AR coefficients.
Rather, as in
\cite{Sasaki2018:Declipping.multiple.matrix.rank.minimization},
we exploit the fact that the Hankel matrix constructed from a block of an AR signal is low-rank.
While \cite{Sasaki2018:Declipping.multiple.matrix.rank.minimization}
treats clipped audio, we propose an optimization problem dealing with missing samples and solve it by an algorithm similar to the referenced one.
Since missing audio usually appears in compacts blocks of samples, in the experiments we
focus on this particular case.

Our first contribution is the introduction of a novel method into the context of audio inpainting.
Second, while the authors in \cite{Sasaki2018:Declipping.multiple.matrix.rank.minimization}
evaluate their respective algorithm solely on speech,
our approach, which includes several proposed modifications, is tested against the state-of-the-art methods
also on a~standard music dataset.

The main idea of \cite{Sasaki2018:Declipping.multiple.matrix.rank.minimization} is actually not brand new.
Optimization involving Hankel matrices has been proposed in
\cite{Takahashi2013:Hankel_matrix_declipping} and 
\cite{Takahashi2015:Block.Adaptive.Declipping.Null.Space}
for the case of audio declipping,
and in \cite{Yokota2022:Soft.smoothness.for.audio.inpainting.using.latent.matrix.model.ARXIV}
for audio inpainting.
Compared to these works,
\cite{Sasaki2018:Declipping.multiple.matrix.rank.minimization} 
proposes to involve \emph{multiple} matrices in the optimization,
leading to an improved reconstruction efficiency.

\section{Method}
\label{sec:method}

The method is based on the assumption that audio signals
$\x = \{x_t\}$
can be modeled as autoregressive (AR). An AR model of order $r$ characterizes the signal samples as depending on $r$ preceding ones,
\begin{equation}
	x_t = \sum_{k=1}^r a_kx_{t-k} + \varepsilon_t,
	\label{eq:AR_process}	
\end{equation}
where $a_k$ are the AR coefficients and $\varepsilon_t$ denotes noise.

Given a signal $\x$, elements of the Hankel-structured matrix $\X\in\RR^{M\times N}$ are defined as
\begin{equation}
	X_{i,j} = x_{i+j-1},
	\label{eq:hankel}
\end{equation}
where $M>N$.
Rows are indexed by $i$ and columns by $j$.
Note that this way, each antidiagonal of $\X$ is constant.
Let $H\subset\RR^{M\times N}$ denote the set (actually a vector space) of Hankel matrices of size $M\times N$.

The key observation is that when noise is ignored in \eqref{eq:AR_process},
the rank of the corresponding Hankel matrix is equal to the order of the AR process, $r$.
This motivates the actual formulation of the inpainting problem, where the rank is being minimized,
i.e., the minimum-order AR model of the signal is searched for, given the constraints.
Let the set $\Gammaset$ summarize the inpainting conditions,
using the reliable samples from the observed signal:
\begin{equation}
	\Gammaset = \left\{ \X \in \RR^{M\times N} \mid \X_{i,j} = \Y_{i,j} \ \text{for}\ i+j-1 \in R \right\},
	\label{eq:gamma_R}
\end{equation}
where $\Y$ represents a Hankel matrix created from the depleted input signal, and $R$ denotes the set of indexes corresponding to the \emph{reliable} audio samples.
The basic optimization problem then reads
\begin{equation}
	\min_{\X}\,\rank(\X) \ \text{s.t.}\  \X\in\Gammaset\cap H,
	\label{eq:basic_problem}
\end{equation}
i.e., a matrix with minimum rank among all matrices satisfying the two feasible conditions is searched for.

Such a problem is NP-hard, and therefore the solution must be only approximated.
Often, the nuclear norm $\| \X \|_*$ is utilized in the literature as a convex surrogate to the nonconvex $\rank(\X)$ function
\cite{Recht2010:nuclear.norm,Candes:2011Robust.princ.comp.anal,DankovaRajmicJirik2015:LVA}.
This makes the problem computationally affordable.
However, the performance of such a~basic form of the reconstruction approach is reported as poor in the context of audio declipping \cite{Sasaki2018:Declipping.multiple.matrix.rank.minimization},
and the same was observed by us in the field of audio inpainting.

To improve the performance, the authors of
\cite{Sasaki2018:Declipping.multiple.matrix.rank.minimization} shift to a~modified problem, called
\emph{multiple} matrix problem:
\begin{equation}
\begin{split}
	\min_{\X,\D_i} &\sum_{i=1}^L \rank(\D_i\X) \\ \text{s.t.}\ &\sum_{i=1}^L\D_i=\I,\ \D_i\in D,\ \X\in\Gammaset\cap H.
	\label{eq:multiple_matrix_problem}
\end{split}
\end{equation}
Here, $L$ is a constant representing the number of matrices $\D_i$, 
$D$~denotes the set
of diagonal matrices whose elements are either 0 or 1, 
and $\mathbf{I}$ represents the identity matrix.
To put it in words,
\eqref{eq:multiple_matrix_problem} splits the processed signal
into $L$ parts, and the rank of each of the parts is minimized separately.
Yet, the parts taken together have to comply with the restrictions given by both $\Gammaset$ and $H$.
The division of each audio sample into one of the $L$ blocks, coded by the diagonal matrices $\D_i$, is optimized jointly with~$\X$.

\begin{algorithm}
\SetAlgoVlined
\DontPrintSemicolon
\SetKwInput{Input}{Input}
\SetKwInput{Init}{Initialization}
\SetKwInput{Par}{Parameters}
	\Input{$\Y, H\cap\Gammaset, \{\D_i\}_{i=1}^L$}
	\Par{$\alpha, \alpha_\text{min}, \eta_\alpha, \lambda, \tau, t_\text{max}$}
	\Init{$t=0, \X=\Y$}
	\Repeat{$t_\textup{max}<t$}{
		$\alpha \leftarrow \max(\alpha/\eta_\alpha, \alpha_\textup{min})$\\
		\For{$i=1,\dots,L$}{
			$[\matr{U}, \sigma_1, \sigma_2, \dots, \sigma_N, \matr{V}] \leftarrow \mathrm{svd}(\D_i\X)$\\
			$r_i\leftarrow \argmin_{\hat{r}} \sigma_{\hat{r}} \ \text{s.t.} \ \sigma_{\hat{r}} \geq \alpha\sigma_1$\\
			$\Z_i \leftarrow \mathcal{T}_{r_i, \lambda\sigma_{r_i}}(\D_i\X)$
		}
		\For{$(i,j)\in\{1,2,\dots,L\}\times\{1,2,\dots,N\}$}{
			$(\d^{(i)})_j \leftarrow \max\left(0, \frac{1}{L}\left(1-\sum_{k=1}^L{\frac{\langle\z^{(k)}_j, \x_j\rangle}{\langle\x_j,\x_j\rangle}}\right) + 
			\frac{\langle\z^{(i)}_j, \x_j\rangle}{\langle\x_j, \x_j\rangle}-\tau \right)$
		}
		$(\d^{(i)})_j \leftarrow (\d^{(i)})_j/\sum_{k=1}^L{(\d^{(k)})_j} \ \forall \ i,j$\\
		$\X \leftarrow \left(\sum_{k=1}^L{\D_i^2}\right)^{-1} \left(\sum_{k=1}^L{\D_i\Z_i}\right)$\\
		$\X \leftarrow \proj_{H\cap\Gammaset}(\X)$\\
		$t\leftarrow t+1$\\
	}
 	\KwRet{$\X$}
	\caption{Inpainting algorithm based on IPMS}
	\label{alg:hankel.inpainting.IPMS}
\end{algorithm}

An approximate numerical solution to problem \eqref{eq:multiple_matrix_problem} can be obtained using the iterative partial matrix shrinkage (IPMS) algorithm \cite{Konishi2014:IPMS.algorithm}.
It is a~heuristic algorithm with roots in proximal splitting
\cite{combettes2011proximal,Condat2019:Proximal.splitting.algorithms}.
The IPMS algorithm can be decomposed into three fundamental steps
corresponding to the requirements of
problem \eqref{eq:multiple_matrix_problem},
see Alg.~\ref{alg:hankel.inpainting.IPMS}.

The first step consists in enforcing the low rank of each $\D_i\X$.
This is achieved by thresholding the singular values of $\D_i\X$.
Ignoring for the moment the index $i$, 
the singular values $\sigma_n$ are obtained via the classic singular value decomposition (SVD):
$[\matr{U}, \sigma_1, \dots, \sigma_N, \matr{V}] = \mathrm{svd}(\D_i\X)$.
While the usual way of processing the singular values would be to apply soft thresholding 
\cite{Donoho1995_De-noising.by.soft.thresholding}
to all of them, the authors of \cite{Sasaki2018:Declipping.multiple.matrix.rank.minimization}
utilize the \emph{partial} soft thresholding operator
$\mathcal{T}_{r, \lambda\sigma_{r}}$\!.
This operator performs the soft thresholding
on the $N-r$ smallest singular values
among $\sigma_1, \dots, \sigma_N$,
with the adaptively derived threshold $\lambda\sigma_{r}$.
The largest $r$ singular values are kept unchanged~\cite{Konishi2014:IPMS.algorithm}.
This way, matrices $\Z_i$ are obtained.

The second step includes an update of $\D_i$.
The preceding thresholding step does not provide matrices $\D_i$ that would obey the constraints $\D_i\in D$.
Therefore, the update of $\D_i$ is done by approximating the solution of the problem
\begin{equation}
	\min_{\hat{\D}_1, \dots, \hat{\D}_L} \sum_{i=1}^L \|\Z_i - \hat{\D}_i\X\|^2_\textup{F}
\label{eq:D_update}
\end{equation}
using a heuristic shrinkage technique, pushing the values of the diagonal elements $(\d^{(i)})_j$ closer to 0 or 1.
The updated matrices $\D_i$ satisfy the condition $\sum_{i=1}^L \D_i = \I$.

Finally, the third principal step involves the update of $\X$ such that it minimizes the distance between $\Z_i$ and $\D_i\X$, 
followed by the projection of $\X$ onto the intersection $H\cap\Gammaset$, which enforces the other two simultaneous feasible conditions.

It is not hard to show that $\proj_{H\cap\Gammaset}$ can be computed as a~composite projection $\proj_{\Gammaset} \circ \proj_H$, 
where the projection onto the Hankel space is evaluated by taking the averages of the antidiagonals, 
while $\proj_{\Gammaset}$
replaces the samples at reliable indexes by the respective audio samples from the observation matrix $\Y$.
Note that opposite to audio declipping, in the case of inpainting the order of the two projections does not matter
but computing the $\proj_H$ first is more computationally effective,
since the consecutive $\proj_{\Gammaset}$ can be done on the signal $\x$, rather than on the corresponding Hankel matrix.
Nonetheless, in the source codes provided, the authors compute only the projection onto the reliable set in each iteration, 
while the projection onto the Hankel space is done only once at the very end of iterations, 
before the Hankel matrix $\X$ is converted back to the time-domain vector.
In the experiments in Sec.~\ref{sec:experiments} we include both variants of the algorithm denoted as \oznaceniMetody{\hankelI{}} and \oznaceniMetody{\hankelHI{}}
to evaluate the influence of $\proj_H$ in each iteration on the performance of the algorithm.

As a final note, the algorithm provided in Alg.~\ref{alg:hankel.inpainting.IPMS} requires an initial setting of $\D_i$.
The idea is that since audio signals do not often switch the AR model,
the matrices $\D_i$ are initialized 
such that the values of $\D_i$ blend smoothly from one AR process to another and that it holds $\sum_{i=1}^L(\d^{(i)})_j = 1$.
The paper \cite{Sasaki2018:Declipping.multiple.matrix.rank.minimization} provides an explicit initialization formula; 
however, the implementation provided by its authors slightly differs.
The above-described character of the initialization is nevertheless preserved.

\section{Block processing}
\label{sec:block_processing}

The original IPMS audio declipping method \cite{Sasaki2018:Declipping.multiple.matrix.rank.minimization} was designed to process the signal by short, overlapping windows.
To avoid breaking the AR signal assumption, the rectangular analysis window is used by the authors;
commonly used non-rectangular windows would lead to weakening the assumption of the deterministic part of \eqref{eq:AR_process}.
However, in the signal synthesis (i.e., after a~processed block has been restored) the authors \emph{replace} the overlapping samples
with the currently processed block.
This may lead to waveform discontinuities in the transitions between blocks, which may cause undesirable artifacts.

To cope with this issue, we propose two approaches to smoothing out the block transitions.
The first approach utilizes crossfading---a commonly-known technique to smoothly progress from one signal segment to another \cite{ZaviskaRajmicMokry2022:Declipping.crossfading}.
The exploited crossfading function is based on the squared sine wave 
\vspace{-4pt}
\begin{equation}
	c_k = \sin^2\left(\frac{k\pi}{2(K+1)}\right), \quad k=1,\dots,K,
\label{eq:xfade}
\end{equation}
where $K$ represents the length of the crossfaded section.

The second approach is based on the standard synthesis overlap-add (OLA) technique.
A~rectangular window is still used for analysis;
however, for the signal synthesis a~smooth window is used.
The currently processed block is weighted by the synthesis window and added to the already-processed part of the signal.
This ensures smooth blending of the currently processed block into the rest of the signal without discontinuities.
For this application we chose the Hann window, which satisfies the important partition-of-unity property in the case of a~75\% overlap.

\section{Experiments and results}
\label{sec:experiments}

This section describes the experiments designed to evaluate the performance of the proposed method and its variants, 
and presents the numerical results of the restoration.
\subsection{Data}
The experiments were performed on a~dataset from the Audio Inpainting Toolbox\footnote{\url{http://small.inria.fr/keyresults/audio-inpainting/}}
accompanying the seminal article on audio inpainting by Adler \emph{et al.} \cite{Adler2012:Audio.inpainting}.
The toolbox contains 10~musical and 10~speech (5~male and 5~female) uncompressed monophonic audio excerpts sampled at 16\,kHz with a duration of 5~seconds
and a~bit-rate of 256~kbps.

To simulate the loss of the time-domain samples, we generated ten gaps randomly distributed along the length of the signals,
and zeroed the audio samples belonging to these intervals.
To ensure a fair evaluation and comparison, positions of the gaps remained fixed for all tested signals and methods.
The performed experiments utilized gaps ranging from 10~ms (160~samples)
up to 50~ms (800~samples) with a~step of 10~ms.

\subsection{Metrics}
As the measure of restoration quality, we use the signal-to-noise ratio (SNR), 
which evaluates the physical similarity of waveforms in decibels such that
\vspace{-2pt}
\begin{equation}
	\mathrm{SNR}(\y, \hat{\y}) = 10 \cdot \log_{10} \frac{\|\y\|^2_2}{\|\y-\hat{\y}\|^2_2},
\label{eq:snr}
\end{equation}
where $\y$ represents the original signal, which is in real situations unknown, and $\hat{\y}$ denotes the restored signal.

Since the physical similarity of waveforms does not necessarily mean the most auditory pleasant result, 
we also use two perceptually motivated metrics---PEMO-Q \cite{Huber:2006a} for music 
and Perceptual Evaluation of Speech Quality (PESQ) \cite{Rix2001:PESQ} for speech audio excerpts.

PEMO-Q has originally  been developed for rating audio quality degraded by compression algorithms; 
nevertheless, it is commonly used also for evaluating the performance of various audio restoration tasks, such as inpainting, declipping, dequantization, etc.
Its output is a number called Objective Difference Grade (ODG), rating the severity of audio degradation in the range from \textminus4 (very annoying) to 0 (imperceptible).

PESQ is a family of standards developed to model the subjective tests commonly used in telecommunications.
It evaluates the quality of speech signal on a Mean Opinion Score (MOS) scale ranging from 1 (bad) to 5 (excellent).

The experiments include four different variants of the proposed inpainting algorithm.
Two of them differ in the projection step---\oznaceniMetody{\hankelI{}} projects only on the feasible set $\Gammaset$, 
while \oznaceniMetody{\hankelHI{}} computes the projection on the intersection of the Hankel space $H$ 
and set of feasible solutions $\Gammaset$ in each iteration.
The other two variants are based on \oznaceniMetody{\hankelHI{}} and utilize an additional step of smoothing the transitions between signal blocks:
\oznaceniMetody{\hankelHIxfade{}} uses crossfade governed by the squared sine wave
\eqref{eq:xfade}
and \oznaceniMetody{\hankelHIola{}} exploits the traditional OLA approach with the Hann window.

The parameters of the proposed method are summarized in Table~\ref{tab:parameters}.
\begin{table}[ht]%
\centering
\caption{Parameters of the IPMS inpainting algorithm}
\begin{tabular}{|l|c|c|}
	\hline
	number of samples in processed block & $K$ & 1024 \\
	number of columns in the Hankel matrix & $N$ & 128 \\
	number of submatrices & $L$ & 3 \\ 
	maximum number of iterations & $t_\text{max}$ & 3000 \\
	heuristic parameter for matrix rank estimation & $\alpha$ & 1 \\
	minimum value of $\alpha$ & $\alpha_\text{min}$ & 0.005 \\
	step size & $\eta_\alpha$ & 1.01 \\
	thresholding coefficient & $\lambda$ & 0.1 \\
	heuristic thresholding parameter & $\tau$ & 0.1 \\
	\hline
\end{tabular}
\label{tab:parameters}
\end{table}

\subsection{Algorithms}
\vspace{-.06em}
The performance of the proposed inpainting algorithms was compared with several top-performing audio inpainting methods; 
namely nonnegative matrix factorization (NMF)-based EM1 \cite{Mokry2022:Audio.inpainting.NMF}, 
the Janssen method \cite{javevr86}, 
interpolation based on the left-sided and right-sided AR-parameter vectors (LR) \cite{Etter1996:Interpolation_AR}, 
the analysis sparse audio inpainter (A-SPAIN) \cite{MokryZaviskaRajmicVesely2019:SPAIN}, 
and an A-SPAIN variant utilizing a~dictionary learning approach
\mbox{(A-SPAIN-learned)}~\cite{TaubockRajbamshiBalasz2021:SPAINMOD}.

The algorithms used in the experiments, including the proposed Hankel-based method,
were implemented and tested in MATLAB 2022a.
Except for the LR method, they all exploit the block-wise processing with 
1024-sample-long overlapping windows and 256 samples window shift (75\% overlap).

Note that the experiments were performed also using the recently introduced NMF-based methods \oznaceniMetody{AM} and \mbox{\oznaceniMetody{AMtoEM1}}~\cite{Mokry2022:Audio.inpainting.NMF}, 
the modified version of the Janssen method~\cite{Mokry2022:Audio.inpainting.NMF}, 
A-SPAIN-mod \cite{TaubockRajbamshiBalasz2021:SPAINMOD}, 
and the weighted variant of the \mbox{$\ell_1$-minimization} approach \cite{MokryRajmic2020:Inpainting.revisited}.
Nevertheless, these methods were omitted from the presented figures for clarity;
their respective results were among neither the best nor the worst.

The source codes of the presented Hankel-based inpainting methods are available by the authors upon request.

\begin{figure*}[t] 
\centering
	\hspace{5em}
	\subfloat[][SNR]{
%
%
\definecolor{mycolor1}{rgb}{0.00000,0.44700,0.74100}%
\definecolor{mycolor2}{rgb}{0.85000,0.32500,0.09800}%
\definecolor{mycolor3}{rgb}{0.92900,0.69400,0.12500}%
\definecolor{mycolor4}{rgb}{0.49400,0.18400,0.55600}%
\definecolor{mycolor5}{rgb}{0.46600,0.67400,0.18800}%
\definecolor{mycolor6}{rgb}{0.30100,0.74500,0.93300}%
\definecolor{mycolor7}{rgb}{0.63500,0.07800,0.18400}%
\begin{tikzpicture}[trim axis left, trim axis right, scale=0.7]

\begin{axis}[%
width=3in,
height=2in,
at={(0.758in,0.481in)},
scale only axis,
axis on top,
xmin=10,
xmax=50,
xtick={10, 20, 30, 40, 50},
xlabel style={font=\color{white!15!black}},
ylabel shift=-5pt,
xlabel={gap length (ms)},
ymin=0,
ymax=10,
ylabel style={font=\color{white!15!black}},
ylabel={SNR (dB)},
axis background/.style={fill=white},
legend style={at={(1.1,1.001)}, anchor=north west, legend cell align=left, align=left, draw=white!15!black}, 
every axis plot/.append style={thick}
]
\addplot [color=mycolor3, dashed]
  table[row sep=crcr]{%
10	9.45468068241232\\
20	6.38355307480676\\
30	4.27102551810812\\
40	2.17806892150137\\
50	1.50881521617655\\
};
\addlegendentry{EM1}



\addplot [color=mycolor4, dashed]
  table[row sep=crcr]{%
10	4.89345419839198\\
20	3.47561809808941\\
30	2.88152440394188\\
40	0.848751106906104\\
50	0.84658225057137\\
};
\addlegendentry{Janssen}


\addplot [color=mycolor5, dashed]
  table[row sep=crcr]{%
10	4.75142367035134\\
20	3.0861639161453\\
30	2.54247889050888\\
40	1.9756991937894\\
50	1.50192217289994\\
};
\addlegendentry{LR}

\addplot [color=mycolor6, dashed]
  table[row sep=crcr]{%
10	6.33530730256312\\
20	3.28151331738251\\
30	1.63088264580179\\
40	0.710882097478448\\
50	0.774064187580383\\
};
\addlegendentry{A-SPAIN}


\addplot [color=mycolor7, dashed]
  table[row sep=crcr]{%
10	7.96341861271969\\
20	6.34479230716576\\
30	5.32945853476205\\
40	3.67818354720611\\
50	3.31975289311079\\
};
\addlegendentry{A-SPAIN-learned}


\addplot [color=mycolor1]
  table[row sep=crcr]{%
10	3.53502964629963\\
20	1.35534191996912\\
30	0.786907906301263\\
40	0.399188431794677\\
50	0.539426508830476\\
};
\addlegendentry{\hankelI{}}

\addplot [color=mycolor2]
  table[row sep=crcr]{%
10	4.4765231795994\\
20	2.89666517519575\\
30	1.97522733604968\\
40	1.12879605956186\\
50	0.764356079343299\\
};
\addlegendentry{\hankelHI{}}

\addplot [color=mycolor3]
  table[row sep=crcr]{%
10	4.8396144126721\\
20	3.2342188393611\\
30	2.06908986783326\\
40	1.17032676298222\\
50	0.78702613947488\\
};
\addlegendentry{\hankelHIxfade{}}

\addplot [color=mycolor4]
  table[row sep=crcr]{%
10	4.76500990472228\\
20	3.3583948293869\\
30	2.23292867525342\\
40	1.22118089013708\\
50	0.986373782046048\\
};
\addlegendentry{\hankelHIola{}}

\end{axis}
\end{tikzpicture}%
\label{subfig:results_music_SNR}}
	\hfill
	\subfloat[][PEMO-Q]{
%
%
\definecolor{mycolor1}{rgb}{0.00000,0.44700,0.74100}%
\definecolor{mycolor2}{rgb}{0.85000,0.32500,0.09800}%
\definecolor{mycolor3}{rgb}{0.92900,0.69400,0.12500}%
\definecolor{mycolor4}{rgb}{0.49400,0.18400,0.55600}%
\definecolor{mycolor5}{rgb}{0.46600,0.67400,0.18800}%
\definecolor{mycolor6}{rgb}{0.30100,0.74500,0.93300}%
\definecolor{mycolor7}{rgb}{0.63500,0.07800,0.18400}%
\begin{tikzpicture}[trim axis left, trim axis right, scale=0.7]

\begin{axis}[%
width=3in,
height=2in,
at={(0.758in,0.481in)},
scale only axis,
axis on top,
xmin=10,
xmax=50,
xtick={10, 20, 30, 40, 50},
xlabel style={font=\color{white!15!black}},
xlabel={gap length (ms)},
ymin=-4,
ymax=-0.5,
ylabel style={font=\color{white!15!black}},
ylabel={PEMO-Q ODG},
axis background/.style={fill=white},
legend style={at={(1.03,1.001)}, anchor=north west, legend cell align=left, align=left, draw=white!15!black}, 
every axis plot/.append style={thick}
]
\addplot [color=mycolor3, dashed]
  table[row sep=crcr]{%
10	-1.01709910232049\\
20	-2.25042065292947\\
30	-3.03432401354695\\
40	-3.50279630478326\\
50	-3.58751790753564\\
};

%

\addplot [color=mycolor4, dashed]
  table[row sep=crcr]{%
10	-0.993858477626274\\
20	-2.13423680425772\\
30	-2.78330185171884\\
40	-3.16477102424479\\
50	-3.34304485601176\\
};


\addplot [color=mycolor5, dashed]
  table[row sep=crcr]{%
10	-1.73285284852904\\
20	-2.94468336717174\\
30	-3.28051887424076\\
40	-3.42397070371511\\
50	-3.49052911170082\\
};

\addplot [color=mycolor6, dashed]
  table[row sep=crcr]{%
10	-0.755610702012853\\
20	-1.79748232267801\\
30	-2.66461199258285\\
40	-3.07022592454522\\
50	-3.32033616825967\\
};


\addplot [color=mycolor7, dashed]
  table[row sep=crcr]{%
10	-1.02561606212719\\
20	-1.97173663963055\\
30	-2.64529235169999\\
40	-2.97444289039717\\
50	-3.13369390039348\\
};


\addplot [color=mycolor1]
  table[row sep=crcr]{%
10	-1.7903842538397\\
20	-3.31248210987477\\
30	-3.56807024445571\\
40	-3.64838596189914\\
50	-3.69086454501031\\
};

\addplot [color=mycolor2]
  table[row sep=crcr]{%
10	-1.38412767844413\\
20	-3.00773186395708\\
30	-3.39536412885277\\
40	-3.56280486553685\\
50	-3.63734464344484\\
};

\addplot [color=mycolor3]
  table[row sep=crcr]{%
10	-1.47120421329651\\
20	-2.98334511307476\\
30	-3.36377888043194\\
40	-3.5441804223777\\
50	-3.62527100538355\\
};

\addplot [color=mycolor4]
  table[row sep=crcr]{%
10	-1.42638183863845\\
20	-2.92882813290249\\
30	-3.3465941277784\\
40	-3.51237729124808\\
50	-3.6113956047927\\
};

\end{axis}
\end{tikzpicture}%
\label{subfig:results_music_PEMO-Q}}
	\hspace{4em}
\vspace{-0.2em}	
\caption{SNR and PEMO-Q audio inpainting results evaluated on the \qm{music} dataset.
}%
\vspace{-1.4em}
\label{fig:results_music}%
\end{figure*}
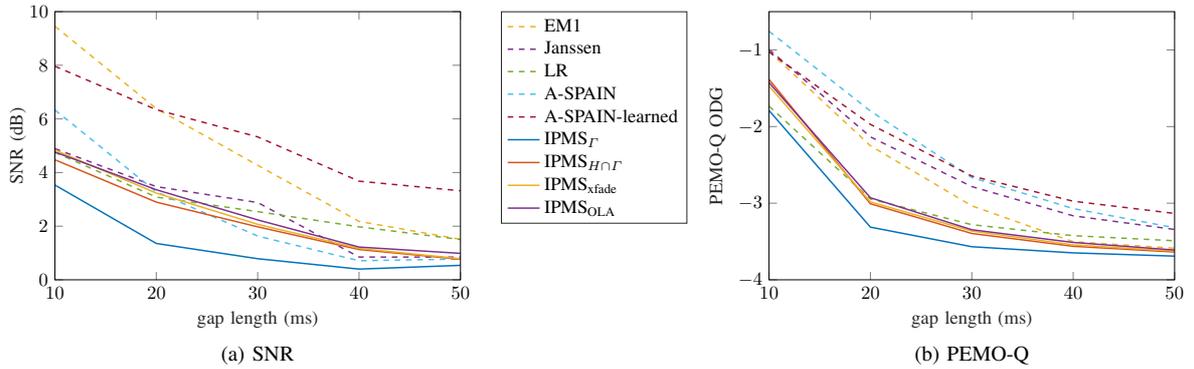

\begin{figure*}[t]
\centering
	\hspace{5em}
	\subfloat[][SNR]{
%
%
\definecolor{mycolor1}{rgb}{0.00000,0.44700,0.74100}%
\definecolor{mycolor2}{rgb}{0.85000,0.32500,0.09800}%
\definecolor{mycolor3}{rgb}{0.92900,0.69400,0.12500}%
\definecolor{mycolor4}{rgb}{0.49400,0.18400,0.55600}%
\definecolor{mycolor5}{rgb}{0.46600,0.67400,0.18800}%
\definecolor{mycolor6}{rgb}{0.30100,0.74500,0.93300}%
\definecolor{mycolor7}{rgb}{0.63500,0.07800,0.18400}%
\begin{tikzpicture}[trim axis left, trim axis right, scale=0.7]

\begin{axis}[%
width=3in,
height=2in,
at={(0.758in,0.481in)},
scale only axis,
axis on top,
xmin=10,
xmax=50,
xtick={10, 20, 30, 40, 50},
xlabel style={font=\color{white!15!black}},
xlabel={gap length (ms)},
ymin=0,
ymax=9,
ylabel style={font=\color{white!15!black}},
ylabel={SNR (dB)},
axis background/.style={fill=white},
legend style={at={(1.1,1.001)}, anchor=north west, legend cell align=left, align=left, draw=white!15!black}, 
every axis plot/.append style={thick}
]
\addplot [color=mycolor3, dashed]
  table[row sep=crcr]{%
10	7.11250458653549\\
20	5.07056591137999\\
30	2.59648992513446\\
40	1.18555237139007\\
50	0.871969529339562\\
};
\addlegendentry{EM1}

%

\addplot [color=mycolor4, dashed]
  table[row sep=crcr]{%
10	8.28549080921533\\
20	6.3795105132087\\
30	3.86848112701542\\
40	2.52708984591597\\
50	1.08781970586603\\
};
\addlegendentry{Janssen}


\addplot [color=mycolor5, dashed]
  table[row sep=crcr]{%
10	5.55557764476263\\
20	3.86804901574935\\
30	2.59196203041147\\
40	1.78487171755909\\
50	1.40339146514975\\
};
\addlegendentry{LR}

\addplot [color=mycolor6, dashed]
  table[row sep=crcr]{%
10	6.73183619109131\\
20	4.66364574449961\\
30	2.3392888020514\\
40	1.31030501900729\\
50	1.27028386859925\\
};
\addlegendentry{A-SPAIN}


\addplot [color=mycolor7, dashed]
  table[row sep=crcr]{%
10	6.52025550172954\\
20	5.33511131195925\\
30	3.72566271126857\\
40	2.16439387952889\\
50	1.37366631501954\\
};
\addlegendentry{A-SPAIN-learned}


\addplot [color=mycolor1]
  table[row sep=crcr]{%
10	5.41662112762914\\
20	2.04677443745364\\
30	1.17302918882679\\
40	0.868206304141472\\
50	0.619683612991286\\
};
\addlegendentry{\hankelI{}}

\addplot [color=mycolor2]
  table[row sep=crcr]{%
10	8.06948953213224\\
20	5.81245557805617\\
30	3.82264494021133\\
40	2.44045740169031\\
50	1.63847924033193\\
};
\addlegendentry{\hankelHI{}}

\addplot [color=mycolor3]
  table[row sep=crcr]{%
10	8.31946048350901\\
20	5.62140416219148\\
30	3.51307439698346\\
40	2.47025424232667\\
50	1.74240127739068\\
};
\addlegendentry{\hankelHIxfade{}}

\addplot [color=mycolor4]
  table[row sep=crcr]{%
10	7.9670763193485\\
20	5.30170306076387\\
30	3.46619083523561\\
40	2.1803037584572\\
50	1.49167091971314\\
};
\addlegendentry{\hankelHIola{}}

\end{axis}
\end{tikzpicture}%
\label{subfig:results_speech_SNR}}\hfill
	\subfloat[][PESQ]{
%
%
\definecolor{mycolor1}{rgb}{0.00000,0.44700,0.74100}%
\definecolor{mycolor2}{rgb}{0.85000,0.32500,0.09800}%
\definecolor{mycolor3}{rgb}{0.92900,0.69400,0.12500}%
\definecolor{mycolor4}{rgb}{0.49400,0.18400,0.55600}%
\definecolor{mycolor5}{rgb}{0.46600,0.67400,0.18800}%
\definecolor{mycolor6}{rgb}{0.30100,0.74500,0.93300}%
\definecolor{mycolor7}{rgb}{0.63500,0.07800,0.18400}%
\begin{tikzpicture}[trim axis left, trim axis right, scale=0.7]

\begin{axis}[%
width=3in,
height=2in,
at={(0.758in,0.481in)},
scale only axis,
axis on top,
xmin=10,
xmax=50,
xtick={10, 20, 30, 40, 50},
xlabel style={font=\color{white!15!black}},
xlabel={gap length (ms)},
ymin=1,
ymax=4.5,
ylabel style={font=\color{white!15!black}},
ylabel={PESQ ODG},
axis background/.style={fill=white},
legend style={at={(1.03,1.001)}, anchor=north west, legend cell align=left, align=left, draw=white!15!black}, 
every axis plot/.append style={thick}
]
\addplot [color=mycolor3, dashed]
  table[row sep=crcr]{%
10	4.18190727233887\\
20	3.51809144020081\\
30	2.56193490028381\\
40	1.90931020975113\\
50	1.64221295118332\\
};

%

\addplot [color=mycolor4, dashed]
  table[row sep=crcr]{%
10	4.25534963607788\\
20	3.73993091583252\\
30	3.13237872123718\\
40	2.72440507411957\\
50	2.18217132091522\\
};


\addplot [color=mycolor5, dashed]
  table[row sep=crcr]{%
10	3.91816778182983\\
20	3.39657402038574\\
30	2.67495665550232\\
40	2.17022700309753\\
50	1.87487028837204\\
};

\addplot [color=mycolor6, dashed]
  table[row sep=crcr]{%
10	3.99831731319428\\
20	3.58172340393066\\
30	3.05178329944611\\
40	2.6391793012619\\
50	2.27966876029968\\
};


\addplot [color=mycolor7, dashed]
  table[row sep=crcr]{%
10	3.85789577960968\\
20	3.50510795116425\\
30	3.12745053768158\\
40	2.66002571582794\\
50	2.2840957403183\\
};


\addplot [color=mycolor1]
  table[row sep=crcr]{%
10	3.94011931419373\\
20	3.0829585313797\\
30	2.21618182659149\\
40	1.74766080379486\\
50	1.47983915805817\\
};

\addplot [color=mycolor2]
  table[row sep=crcr]{%
10	4.23182244300842\\
20	3.64806597232819\\
30	2.67472794055939\\
40	2.08304694890976\\
50	1.77055262327194\\
};

\addplot [color=mycolor3]
  table[row sep=crcr]{%
10	4.27007780075073\\
20	3.65270648002625\\
30	2.65466854572296\\
40	2.05725628137589\\
50	1.77121788263321\\
};

\addplot [color=mycolor4]
  table[row sep=crcr]{%
10	4.25110540390015\\
20	3.67871470451355\\
30	2.6492390871048\\
40	2.06912339925766\\
50	1.75599956512451\\
};

\end{axis}
\end{tikzpicture}%
\label{subfig:results_speech_PESQ}}
	\hspace{4em}
\vspace{-0.2em}	
\caption{SNR and PESQ audio inpainting results evaluated on the \qm{speech} dataset.%
}%
\vspace{-1em}
\label{fig:results_speech}%
\end{figure*}
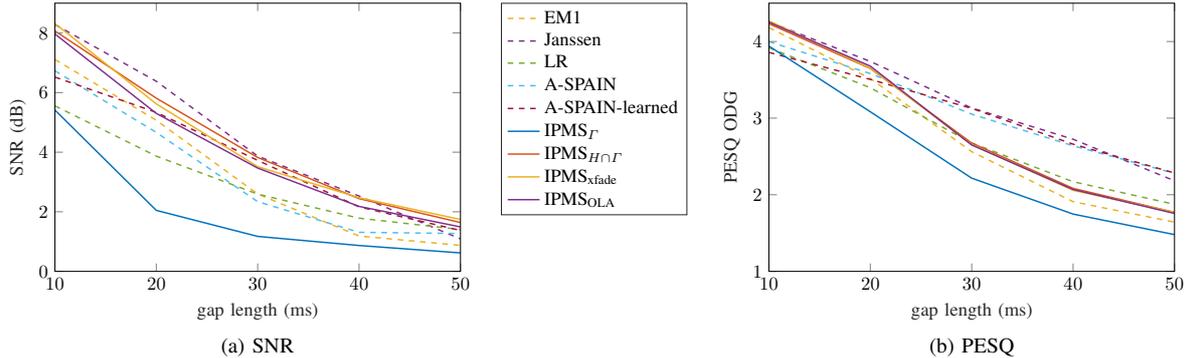

\subsection{Results}
\vspace{-.1em}
The results for the music dataset are illustrated in Fig.~\ref{fig:results_music}.
The results show that performing the projection onto the Hankel space significantly improves the inpainting results according to both the SNR and PEMO-Q.
Smoothing the transitions between the individual blocks of signal using crossfading or the OLA increases the performance even more,
although in these cases the difference is less significant than in the case of the Hankel space projection.
Still, the overall results for music signals indicate that the proposed method is not competitive with the current state-of-the-art methods.

Nonetheless, the situation is different for speech signals.
The results in Fig.~\ref{fig:results_speech} show that while \oznaceniMetody{\hankelI{}} loses to other methods, 
other variants of IPMS provide very good SNR results that are comparable to the Janssen method.
Surprisingly, the best option for speech seems to be the plain \oznaceniMetody{\hankelHI{}} \emph{without} transition smoothing.
The PESQ results are, however, more critical and show that while the proposed method outperforms all but one methods for shorter gaps (10 and 20~ms), 
its performance drops significantly for gaps that are longer.

We offer an explanation why the algorithm performed better on speech than on music:
The speech of a~single speaker can be efficiently modeled exploiting an AR process,
a~fact that is utilized in speech coders, for instance.
On the other hand, AR modeling of musical instruments is in general not that efficient.
On top of that, when polyphonic/multiinstrument pieces are considered, 
the signal is formed as a~sum of components.
These can be AR-modelable separately,
but the AR properties of individual components do not transfer to their sum.
Loosely speaking, a~sum of AR processes is not an AR process.

Let us comment on the weights $\D_i$, $i=1,\dots,L$ of the sub-processes.
As described at the end of Sec.\,\ref{sec:method}, these weights are initialized
by smooth windows, which seems to be a~reasonable choice, allowing \qm{switching} from one AR process to another, adaptively to the signal contents.
Our observation is that at convergence, the profiles of $(\d^{(i)})_j$ do not form compact groups.
This is surprising. to say the least.
It means that the optimization of \eqref{eq:multiple_matrix_problem} via the IPMS algorithm
leads to a~more or less random partition of the processed blocks into sub-blocks, which in a~sense contradicts the basic AR modeling idea.

The average computational complexity of the \hankelI{} algorithm is summarized in Table \ref{tab:time}.

\begin{table}[t]%
\centering
\caption{Average computation times per a~second of audio}
\begin{tabular}{l|r|r|r|r|r}
	\hline
	gap size (ms) &  10 &  20 &  30 &  40 &  50\\
	\hline
	time (s)      & 159 & 199 & 225 & 262 & 306\\
	\hline
\end{tabular}
\vspace*{-6pt}
\label{tab:time}
\end{table}

\section{Conclusion}
\label{sec:conclusion}

This paper presented a novel audio inpainting algorithm based on Hankel-structured matrix rank minimization,
formerly applied to the problem of audio declipping.
The results of the experiments have shown that an extra projection onto the Hankel space in each iteration significantly improves the performance. 
The algorithm turned out to perform quite well for speech signals but not that well for music signals.

Furthermore, we proposed and examined two possibilities of smoothing the transitions between individual signal blocks---one based 
on time-domain crossfading and the other on the (related) overlap-add approach.
These techniques slightly improved the results for the music signals;
however, in the case of speech signals they did not improve the performance according to PESQ and even slightly worsened the results according to the SNR.

The results highlighted the potential of the proposed algorithm, especially in the case of speech inpainting.
However, the results obtained did not meet the expectations
stemming from the audio declipping performance reported in \cite{Sasaki2018:Declipping.multiple.matrix.rank.minimization}.






%

{
\bibliographystyle{IEEEtran}
\inputencoding{cp1250}
\newcommand{\noopsort}[1]{} \newcommand{\printfirst}[2]{#1}
  \newcommand{\singleletter}[1]{#1} \newcommand{\switchargs}[2]{#2#1}

}


\end{document}